\documentclass[conference]{IEEEtran}\usepackage[font=footnotesize]{caption}
\IEEEoverridecommandlockouts
\usepackage{cite}
\usepackage{amsmath,amssymb,amsfonts}
\usepackage{algorithmic}
\usepackage{graphicx}
\usepackage{textcomp}
\usepackage{xcolor}
\usepackage{caption}

\usepackage{tabularx} 
\usepackage{booktabs}

\usepackage{siunitx}

\usepackage{amsmath}
\usepackage{makecell}

\usepackage[colorlinks=true, allcolors=blue]{hyperref}

\renewcommand\IEEEkeywordsname{Keywords}

\usepackage{array}
\newcommand*{\mline}[1]{%
\begingroup
    \renewcommand*{\arraystretch}{1.1}%
   \begin{tabular}[c]{@{}>{\raggedright\arraybackslash}p{3cm}@{}}#1\end{tabular}%
  \endgroup
}

\def\BibTeX{{\rm B\kern-.05em{\sc i\kern-.025em b}\kern-.08em
    T\kern-.1667em\lower.7ex\hbox{E}\kern-.125emX}}
\begin{document}

\title{Food Data in the Semantic Web: A Review of Nutritional Resources, Knowledge Graphs, and Emerging Applications}

\author{\IEEEauthorblockN{1\textsuperscript{st} Darko Sasanski}
\IEEEauthorblockA{\textit{Faculty of Computer Science and Engineering} \\
\textit{Ss. Cyril and Methodius University}\\
Skopje, North Macedonia \\
darko.sasanski@finki.ukim.mk}
\and
\IEEEauthorblockN{2\textsuperscript{nd} Riste Stojanov}
\IEEEauthorblockA{\textit{Faculty of Computer Science and Engineering} \\
\textit{Ss. Cyril and Methodius University}\\
Skopje, North Macedonia \\
riste.stojanov@finki.ukim.mk}
}

\maketitle

\begin{abstract}
This comprehensive review explores food data in the Semantic Web, highlighting key nutritional resources, knowledge graphs, and emerging applications in the food domain. It examines prominent food data resources such as USDA, FoodOn, FooDB, and Recipe1M+, emphasizing their contributions to nutritional data representation. Special focus is given to food entity linking and recognition techniques, which enable integration of heterogeneous food data sources into cohesive semantic resources. The review further discusses food knowledge graphs, their role in semantic interoperability, data enrichment, and knowledge extraction, and their applications in personalized nutrition, ingredient substitution, food-drug and food-disease interactions, and interdisciplinary research. By synthesizing current advancements and identifying challenges, this work provides insights to guide future developments in leveraging semantic technologies for the food domain.
\end{abstract}


\section{Introduction}
The vast and complex landscape of food-related data presents both challenges and opportunities in the domains of personalized nutrition, health research, and food safety. With the rise of semantic web technologies, food knowledge graphs (KGs) have emerged as powerful tools to organize, integrate, and make sense of diverse food data sources. These graphs enable the effective linking of heterogeneous datasets, including nutritional databases, recipes, food ontologies, and food science research, providing a comprehensive framework for understanding the relationships between food entities.

Key resources such as the USDA National Nutrient Database~\cite{haytowitz2011usda}, FoodOn ontology~\cite{dooley2018foodon}, FooDB~\cite{wishart2018foodb}, and Recipe1M+~\cite{marin2021recipe1m+} have laid the foundation for the development of food knowledge graphs. These resources provide structured and standardized representations of food items, ingredients, and their nutritional properties, which can be leveraged to address a wide range of challenges in the food domain. The integration of these diverse datasets is crucial for applications such as personalized nutrition, ingredient substitution, food-disease interactions, and food safety.

Central to the development of food knowledge graphs is the use of food entity recognition (NER) and linking (NEL) techniques. These methods enable identification and connection of food-related data across multiple datasets, ensuring consistency and enhancing data interoperability. Such techniques are essential for creating more accurate and reliable systems that can inform personalized nutrition strategies, improve ingredient substitution recommendations, and support interdisciplinary research in the food-health domain.

This paper is structured as follows. The first section examines key food data resources, including USDA, FoodOn, FooDB, and Recipe1M+, providing a foundation for understanding the food data landscape. The second section explores food entity linking and recognition techniques, which enable the integration of heterogeneous data sources into unified semantic structures. The third section focuses on food semantic resources and knowledge graphs, emphasizing their role in enhancing data interoperability and knowledge extraction. The fourth section discusses various applications, such as personalized nutrition, ingredient substitution, and food-drug interactions, demonstrating the impact of these semantic resources. Finally, the conclusion synthesizes key insights, highlights current challenges, and outlines future directions for advancing semantic technologies in the food domain.

\section{Data Resources in the Food Domain}

This section provides an overview of the primary data resources in the food domain, highlighting their key characteristics, scope, and relevance to food science, nutrition, and related research areas. These resources serve as essential building blocks for developing semantic representations, enabling comprehensive analysis and integration of food-related data across various disciplines.

\textbf{Recipe1M+} is a large-scale dataset containing over one million cooking recipes paired with approximately 13 million food images, introduced in~\cite{marin2021recipe1m+}. Each recipe includes structured data such as titles, ingredients (with quantities and units), step-by-step instructions, and metadata like cooking times and nutritional information. This rich dataset has become a foundational resource in computational gastronomy, enabling cross-modal learning, image-recipe retrieval, and various NLP tasks such as named entity recognition and ingredient substitution, fostering advancements in food informatics and semantic data integration.

\textbf{USDA National Nutrient Database} is a comprehensive database developed by the United States Department of Agriculture, consolidating several food composition resources like the Food and Nutrient Database for Dietary Studies (FNDDS) and the USDA Branded Food Products Database. Launched in 2019, it provides detailed nutrient profiles for a wide range of foods, from raw ingredients to branded products. With categories like Foundation Foods, Experimental Foods, and Standard Reference, FoodData Central supports applications in dietary assessment, public health, and nutritional research, making it an essential resource for professionals and researchers in food and nutrition~\cite{haytowitz2011usda}.

In addition to the USDA's FoodData Central, numerous countries have developed their own Food Composition Databases (FCDBs) to support national dietary assessments and public health initiatives. For instance, the European Food Information Resource (EuroFIR) provides access to food composition data from multiple European countries, enabling cross-border nutritional research and policy development~\cite{greenfield2003food, westenbrink2009food}.

\textbf{FoodOn} is an open-source ontology developed to standardize food-related data across various domains, including agriculture, nutrition, and food safety. It provides a controlled vocabulary for naming parts of animals, plants, fungi, and derived food products, enabling data integration and interoperability. By reusing terms from other ontologies, such as environmental terms from ENVO and agricultural terms from AGRO, FoodOn supports FAIR data annotation and sharing objectives across diverse sectors. This harmonization enhances global food traceability, quality control, and data integration, making it a vital resource for research and applications in food informatics~\cite{dooley2018foodon}.

\textbf{FooDB} is an open-access database providing comprehensive information on the chemical constituents of foods, including both macronutrients and micronutrients. It contains data on over 28,000 chemicals found in approximately 1,000 raw or unprocessed food products. Each chemical entry includes detailed compositional, biochemical, and physiological information, such as nomenclature, structure, chemical class, physico-chemical properties, food sources, color, aroma, taste, physiological effects, and concentrations in foods. This resource is invaluable for researchers, dietitians, food scientists, and educators, offering a wealth of information on the chemical makeup of foods and their implications for human health~\cite{wishart2018foodb}.

\textbf{SNOMED CT} (Systematized Nomenclature of Medicine - Clinical Terms) is a comprehensive clinical terminology system that includes codes for various health-related concepts, including food-related conditions. For example, it encompasses codes for food allergies and intolerances and various types of foods~\cite{donnelly2006snomed}.

\textbf{The Hansard corpus} refers to the official verbatim report of debates in the UK Parliament, which often includes discussions on food-related policies and issues. Researchers have utilized the Hansard taxonomy to annotate food entities in various corpora, such as the FoodBase corpus, which contains annotated food entities extracted from recipes~\cite{hansard_corpus, popovski2019foodbase}.

\textbf{FooDis} is a crucial resource developed for extracting food-disease interactions from biomedical literature using advanced natural language processing techniques. By identifying and validating relationships between foods and health conditions, FooDis enables researchers to uncover potential cause-and-effect links that were previously difficult to detect. Its high precision in matching known associations makes it an essential tool for advancing our understanding of how diet influences disease and for enabling nutritional research and healthcare applications~\cite{cenikj2023foodis}.

\textbf{DrugBank} is a comprehensive database that contains detailed information on over 15,000 drugs, including their chemical properties, mechanisms of action, and therapeutic uses. Although its primary focus is on pharmacological data, DrugBank also includes information on food-drug interactions, highlighting how certain foods can influence drug metabolism, absorption, and efficacy. This data is particularly valuable for research in nutritional genomics and personalized medicine, where understanding the interplay between food and drug interactions can improve treatment outcomes and patient safety~\cite{knox2024drugbank}.

\textbf{AllergenOnline} is a peer-reviewed, curated database designed to identify proteins that may present a potential risk of allergenic cross-reactivity. It serves as a valuable tool for assessing the safety of proteins introduced into foods through genetic engineering or processing methods~\cite{goodman2016allergenonline}.

\textbf{The Cooking Ontology} is a structured framework designed to represent key concepts in the culinary domain, including actions, food items, recipes, and utensils. It serves as a foundational resource for enhancing natural language processing capabilities in culinary applications~\cite{batista2006ontology}.

The datasets and resources discussed throughout this section, summarized in Table~\ref{tab:food-datasets}, provide essential data for advancing research in the food domain. These resources play a critical role in various applications, such as personalized nutrition and food composition analysis. To utilize these datasets effectively, it is vital to perform accurate entity linking, ensuring the correct identification and alignment of food-related entities across multiple resources. The next section delves into the methodology of entity linking and recognition in the food domain.

\begin{table*}[ht]
\centering
\caption{Summary of Key Food Datasets and Resources}
\begin{tabular}{|p{3cm}|p{6cm}|p{6cm}|}
\hline
\textbf{Dataset/Resource} & \textbf{Key Features} & \textbf{Primary Application} \\ \hline
\textbf{Recipe1M+} & 1M+ recipes, 13M images, structured data & Cross-modal learning, ingredient substitution \\ \hline
\textbf{USDA National Nutrient Database} & Nutrient profiles for foods, food categories & Dietary assessment, public health, nutritional research \\ \hline
\textbf{FoodOn} & Controlled vocabulary, FAIR data sharing & Food traceability, quality control, data integration \\ \hline
\textbf{FooDB} & Chemical consituents of foods, 28,000+ chemicals, biochemical data & Food composition research \\ \hline
\textbf{SNOMED CT} & Health-related food terms, allergy data & Health research, food allergies \\ \hline
\textbf{The Hansard corpus} & UK Parliament debates, food policies & Food entity annotation, policy research \\ \hline
\textbf{FooDis} & Food-disease interactions & Nutritional research, healthcare \\ \hline
\textbf{DrugBank} & 15,000+ drugs, food-drug interactions & Nutritional genomics, personalized medicine \\ \hline
\textbf{AllergenOnline} & Peer-reviewed allergenic protein database & Allergen research, food safety \\ \hline
\textbf{The Cooking Ontology} & Structured framework for culinary domain & Culinary applications, NLP enhancement \\ \hline
\end{tabular}
\label{tab:food-datasets}
\end{table*}

\section{Food Entity Linking and Recognition}

Entity Linking (EL) and Entity Resolution (ER) are essential tasks in Natural Language Processing (NLP) that involve identifying, disambiguating, and mapping entities in text to a standardized set of identifiers or canonical forms. EL connects mentions of entities to their corresponding entries in a knowledge base, while ER merges different mentions referring to the same real-world entity. Both processes enhance data integration, information extraction, and semantic understanding of textual data~\cite{ibm:entity_resolution, sasanski2025aligning}.

Over the past decade, food Named Entity Recognition (NER) and Named Entity Linking (NEL) have evolved significantly, transitioning from rule-based methods to advanced machine learning techniques.

Early systems focused on rule-based methods to extract food-related information from various sources:

\begin{itemize}
    \item \textbf{drNER}: A rule-based method designed to extract dietary information from evidence-based recommendations, enabling structured analysis of nutritional guidelines~\cite{eftimov2017rule}.

    \item \textbf{FoodIE}: A rule-based method for extracting food entities from unstructured data, utilizing computational linguistics and semantic information~\cite{popovski2019foodie}.

    \item \textbf{StandFood}: A classification method emphasizing the lexical similarity of food entities, linking them to the FoodEx2 database by the European Food Safety Authority (EFSA)~\cite{eftimov2017standfood}.
\end{itemize}

The introduction of the FoodBase annotated corpus~\cite{popovski2019foodbase} marked a key advancement, enabling the use of machine learning techniques:

\begin{itemize}
    \item \textbf{BuTTER}: A machine learning-based method using a Bidirectional Long Short-Term Memory (BiLSTM) network combined with Conditional Random Fields (CRF) for identifying food entities~\cite{cenikj2020butter}.

    \item \textbf{FoodNER}: A set of models fine-tuned from pretrained BERT architectures to extract and annotate food entities in diverse contexts, representing a major step forward in food information extraction~\cite{stojanov2021fine}. 
\end{itemize}

To assist experts in navigating different food standards and improving interoperability, \textbf{FoodViz} was developed as a web-based framework for presenting food annotation results from NLP and machine learning pipelines~\cite{stojanov2020foodviz}.

The study "Zero-shot evaluation of ChatGPT for food named-entity recognition and linking" evaluates the performance of ChatGPT-3.5 and ChatGPT-4 in identifying and linking food-related entities without prior task-specific training. Results show both models perform better with consumption data than scientific literature, with ChatGPT-4 showing slight improvement over ChatGPT-3.5. However, both models struggle with linking entities to standardized identifiers, indicating the need for further refinement~\cite{ogrinc2024zero}.

Despite significant advancements in food NER and NEL, challenges such as variability in food terminology, the scarcity of comprehensive annotated datasets, and the complexity of food compositions remain. Addressing these issues is essential for improving the accuracy and reliability of food information extraction systems~\cite{ogrinc2024zero, akujuobi2024revisiting}. 

Beyond entity recognition and linking, integrating data from heterogeneous food sources introduces additional layers of complexity. For instance, aligning resources like USDA and FooDB is non-trivial: while both offer structured nutritional information, they differ significantly in terminology, unit representation (e.g., mg vs. g), ingredient granularity (e.g., milk vs. whole cow's milk), and conceptual hierarchies. Ontologies such as FoodOn and SNOMED CT further complicate alignment due to overlapping, yet semantically nuanced, definitions. These differences highlight the need for sophisticated linking approaches that go beyond lexical similarity, underscoring the importance of harmonization frameworks and robust evaluation techniques.

\section{Food Semantic Resources and Knowledge Graphs}

Food semantic resources and knowledge graphs integrate diverse food-related data, including nutritional, molecular, and culinary information. By linking datasets from food ontologies, nutritional databases, and recipes, they enable comprehensive analysis and interdisciplinary research in personalized nutrition, food safety, and sustainability. This section overviews key semantic resources and knowledge graphs in the food domain, highlighting their contributions and impact on food science and technology.

\textbf{FoodOntoMap} is a resource developed to link food concepts across various food ontologies, enabling normalization and integration of food-related data. By mapping food concepts extracted from recipes to semantic tags from multiple food ontologies, it enables a unified representation of food information across different domains~\cite{popovski2019foodontomap}.

The dataset includes food concepts extracted from 22,000 recipes sourced from AllRecipes~\cite{groves2013allrecipes}. Each ingredient is annotated using ontologies available in BioPortal~\cite{noy2009bioportal}, such as FoodOn, OntoFood, SNOMED CT, and the Hansard Corpus. This mapping links distinct food ontologies, enabling applications that analyze relationships between food systems, human health, and the environment.

FoodOntoMap addresses the challenge of integrating diverse food ontologies, each developed for specific applications, by providing a unified framework. This interoperability supports the development of tools for studying food-health-environment interactions and enhances the consistency of food data representation.

The resource has been further expanded with additional ontologies, broadening its coverage and reinforcing its role in semantic food data integration. By improving interoperability, FoodOntoMap advances research and application development in nutrition science, food safety, and sustainability.

In summary, FoodOntoMap serves as a crucial tool for linking and normalizing food concepts across ontologies, enabling a more comprehensive understanding of food-related data.

\textbf{FoodKG} is a comprehensive knowledge graph designed to integrate diverse food-related data into a unified structure, connecting food concepts and their relationships across various domains. It provides a semantic framework for understanding food systems, human health, and the environment by linking food entities to a wide range of data, including nutritional information, recipes, and ontologies~\cite{haussmann2019foodkg}.

The construction of FoodKG integrates multiple data sources, such as recipe data from Recipe1M+, nutrient information from the USDA, and semantic resources like FoodOn, forming a cohesive knowledge graph. This integration addresses challenges such as standardizing food data, merging fragmented sources, and enhancing data accessibility. As a result, FoodKG supports applications in personalized nutrition, recipe recommendation, and food-disease research.

FoodKG has enabled the development of practical tools, including a SPARQL-based service that recommends recipes based on available ingredients and constraints like allergies, as well as a cognitive agent for natural language question answering on the knowledge graph.

In summary, FoodKG is an essential resource for connecting and integrating food-related data, improving interoperability, and enabling applications that enhance dietary choices, health outcomes, and environmental sustainability.

\begin{table}[ht!]
\centering
\caption{Inconsistent Mappings for Cheese Ingredients~\cite{sasanski2025aligning}}
\begin{tabular}{|p{4cm}|p{4cm}|}
    \hline
    \textbf{Ingredient} & \textbf{USDA Class} \\
    \hline
    1\% fat cottage cheese & Blue Cheese \\
    \hline
    2\% cheddar cheese & Blue Cheese \\
    \hline
    Anejo cheese & Blue Cheese \\
    \hline
    Better Than Cream Cheese cream cheese substitute & Blue Cheese \\
    \hline
    Cotija cheese & Blue Cheese \\
    \hline
    Greek feta cheese & Blue Cheese \\
    \hline
\end{tabular}
\label{tab:cheese_mappings}
\end{table}

\begin{table}[ht!]
    \centering
    \caption{Linkage Statistics~\cite{sasanski2025aligning}}
    \begin{tabular}{|p{3.7cm}|c|c|}
    \hline
    \textbf{Statistics} & \textbf{FoodKG Count} & \textbf{New KG Count} \\
    \hline
    \multicolumn{3}{|c|}{\textbf{Dataset Linkages}} \\
    \hline
    Number of links to USDA & 5934 & 10417 \\
    \hline
    Number of links to FoodOn & 2192 & 10586 \\
    \hline
    Number of links to FooDB & 0 & 9377 \\
    \hline
    \multicolumn{3}{|c|}{\textbf{Additional Linkage Stats}} \\
    \hline
    \mline{Percentage of ingredients linked to USDA} & \makecell[c]{32.51\%} & \makecell[c]{90.87\%} \\
    \hline
    \mline{Percentage of ingredients linked to FoodOn} & \multicolumn{1}{c|}{12.01\%} & \multicolumn{1}{c|}{92.35\%} \\
    \hline
    \mline{Percentage of ingredients linked to FooDB} & \multicolumn{1}{c|}{0\%} & \multicolumn{1}{c|}{81.80\%} \\
    \hline
    \end{tabular}
    \label{tab:kg_statistics}
\end{table}

In recent advancements, the preprocessing steps for entity resolution (ER) outlined in the FoodKG framework have undergone significant refinement~\cite{sasanski2025aligning}. Initially, the mappings within the knowledge graph were found to be unreliable, which hindered the accuracy and effectiveness of any subsequent research or analysis, as evidenced in Table~\ref{tab:cheese_mappings}. This issue highlighted the need for a more robust and accurate alignment process. As part of the effort to address these limitations, this study~\cite{sasanski2025aligning} has not only reworked the alignment methodology but also significantly expanded the knowledge graph to incorporate critical information on the chemical compositions and health effects of food constituents by integrating the FooDB dataset. This addition enables a richer understanding of food data, linking nutritional and chemical information with health outcomes, an essential aspect for advancing both food science and health research. The comparison of linkage statistics between the original FoodKG and the newly expanded knowledge graph is presented in Table~\ref{tab:kg_statistics}, demonstrating substantial improvements in coverage and data integration.

\begin{table}[ht!]
    \centering
    \caption{Comparison of the mappings~\cite{sasanski2025aligning}}
    \begin{tabular}{|l|c|c|}
        \hline
         & \textbf{Our Mapping} & \textbf{FoodKG's Mapping} \\
        \hline
        \textbf{True Positives} & 1985 & 1078 \\
        \hline
        \textbf{False Positives} & 618 & 1525 \\
        \hline
        \textbf{Precision (\%)} & 76.25 & 41.41 \\
        \hline
    \end{tabular}
    \label{tab:eval_table}
\end{table}

To ensure the reliability of the newly generated mappings, a rigorous evaluation process was implemented. Students annotated a selected subset of these mappings, who provided detailed feedback on the quality and relevance of the links. As shown in Table~\ref{tab:eval_table}, the feedback strongly favored the newly generated mappings, indicating a significant improvement in both accuracy and utility.

The integration of multiple aspects of the food domain, including nutritional information, chemical compositions, and health effects, into a unified knowledge graph represents a significant milestone. For the first time at this scale, these diverse aspects of food data are encoded into a single resource, enabling comprehensive interdisciplinary research. This work lays the foundation for advanced applications in personalized nutrition, disease prevention, and sustainable food systems, providing researchers with a robust tool for a wide range of studies.

\section{Applications of food semantic resources}

Food semantic resources, such as knowledge graphs and linked datasets, have become essential tools in addressing complex challenges in food science, nutrition, and related fields. By integrating diverse data sources, these resources enable advanced applications ranging from personalized nutrition and dietary analysis to food safety, sustainability, and health research. This section explores key applications that leverage food semantic resources to enhance research, improve decision-making, and foster interdisciplinary collaboration.

A basic use case for food semantic resources is their application in personalized recipe recommendation systems. By leveraging SPARQL queries, these systems can filter recipes based on dietary preferences and constraints. For example, users can search for recipes that include or exclude specific ingredients, such as vegetables, meat, or allergens. Nutritional filtering can also be applied, enabling users to find recipes low in sugar, fat, or calories, which is particularly useful for managing health conditions like diabetes or hypertension.

These use cases are demonstrated in the FoodKG paper, where SPARQL queries were employed to filter recipes by ingredients and nutritional values, addressing health-related dietary restrictions. In addition to ingredient and nutritional filtering, these systems can support dietary preferences by categorizing ingredients based on food types, such as vegan, vegetarian, or halal, ensuring recipes align with users' lifestyle choices. The integration of these use cases within food knowledge graphs enables personalized, health-conscious recipe recommendations, helping users make informed dietary choices.

Two approaches highlight the potential of using natural language for querying semantic resources. The first, from the FoodKG paper, demonstrates Knowledge Base Question Answering (KBQA) applied to food-related knowledge graphs. Users can pose natural language questions, and the system retrieves answers from the graph, simplifying interaction with complex data. The paper also introduced a synthetic Q\&A dataset to test this, with questions ranging from simple ones like “How much sugar is in cheese, cream, fat free?” to more complex constraints-based queries like 'What Laotian dishes can I make with sugar, water, oranges?' This method enhances the accessibility of food knowledge graphs, enabling intuitive querying based on natural language.

The second approach, in Bridging the Gap: Generating a Comprehensive Biomedical Knowledge Graph Question Answering Dataset, presents a similar method applied to the biomedical domain. They created a Q\&A dataset based on the PrimeKG graph, offering over 83,000 question-answer pairs that could be easily reproduced in the food domain. Both methods showcase how natural language can make complex knowledge graphs more accessible, enabling intuitive and efficient querying across different domains and highlighting the significant benefit of simplifying the exploration of large, complex datasets~\cite{yan2024bridging}.

A key application of food knowledge graphs is ingredient substitution, which is essential for meeting dietary or health needs. The paper Identifying Ingredient Substitutions Using a Knowledge Graph of Food demonstrates how food knowledge graphs, combined with semantic data and word embeddings, can identify suitable ingredient substitutes based on factors like nutritional content and food classifications~\cite{shirai2021identifying}.

Food knowledge graphs enable automatic generation of ingredient alternatives to accommodate dietary restrictions such as allergies, low sugar diets, or vegan preferences. The system ranks substitutions based on their compatibility with the original recipe, offering flexibility for users to adapt meals to their health goals or ingredient availability. This approach allows individuals to modify familiar meals rather than following rigid meal plans, promoting personalized nutrition.

Despite the potential, current knowledge graphs, like FoodKG, may overlook certain aspects of ingredients, such as their role in recipes, taste, or texture, which can impact substitution accuracy. However, the newly generated semantic resource from~\cite{sasanski2025aligning}, which includes information about the molecular composition of ingredients, could enhance substitution accuracy. Since ingredients with similar molecular structures are likely to share comparable taste and texture, as discussed in~\cite{ahn2011flavor}, this resource has the potential to suggest substitutions that preserve the original recipe's flavor and structure. This approach would improve the accuracy of ingredient recommendations, ensuring that dietary restrictions are met while maintaining the sensory qualities of the original dish.

Integrating the FoodOn ontology into food knowledge graphs enhances various applications by categorizing food entities based on their origin, processing, and supply chain relationships~\cite{dooley2018foodon}. This improves food traceability, enabling identification of contamination sources and affected products, recipes, or distribution points, thereby reducing economic losses and health risks.

Integrating FoodOn with species and anatomical ontologies enables identification of toxic compounds and analysis of interactions between food components and medications. By linking food chemicals to the Comparative Toxicogenomics Database (CTD), researchers can assess how dietary compounds influence gene behavior and contribute to disease risk. This approach not only strengthens food safety measures but also helps tailor dietary recommendations to individual health needs~\cite{dooley2018foodon, davis2017comparative, wishart2006drugbank}.

Additionally, aligning with resources like FooDB and FoodOn helps establish food-disease relationships, enabling personalized recipe recommendations tailored to specific health needs~\cite{cenikj2023foodis}. In summary, integrating FoodOn strengthens food traceability, safety, and personalized nutrition through a standardized framework for analyzing food entities and their properties.

\section{Conclusion}
In this review, we explored the evolving landscape of food data and the pivotal role of semantic technologies, particularly food knowledge graphs—in transforming how food-related information is organized, integrated, and applied. We examined key resources such as USDA, FoodOn, FooDB, and Recipe1M+, highlighting their foundational roles in food knowledge representation. We also reviewed food entity recognition and linking techniques, emphasizing how they enhance interoperability and enable applications in personalized nutrition, ingredient substitution, and food safety.

While the integration of these resources is already driving innovations in personalized nutrition, improving health outcomes, and enabling interdisciplinary research in the food-health domain. However, challenges remain, including the need for more comprehensive datasets, standardization, and effective methods for handling data inconsistencies.

Unlike prior reviews that focus on isolated datasets or applications, this work offers a comprehensive synthesis across nutritional databases, ontologies, and knowledge graphs. It is, to our knowledge, the first to jointly analyze food entity recognition techniques, dataset alignment strategies, and semantic applications in food science. This integrated perspective also highlights the emerging role of large language models (LLMs) in supporting semantic alignment and content enrichment.

Looking ahead, the continued development of food knowledge graphs and the refinement of entity recognition and linking techniques will further enhance the accuracy and utility of these resources. Future research should focus on expanding the scope of food knowledge graphs, incorporating more diverse food data, and improving the scalability of these systems for broader adoption. By doing so, we can unlock even greater potential for personalized health interventions, enhanced food safety, and more sustainable food systems.

\bibliography{cas-ref}

\end{document}